\documentclass[dvips]{article}
\usepackage{graphicx}
\usepackage{amssymb}
\usepackage{graphicx}
\usepackage{dcolumn}
\usepackage{amsmath}
\usepackage{lscape}
\usepackage{longtable}
\usepackage{subfigure}
\usepackage{color}

\begin{document}
\newcommand{\bd}{\begin{document}}
\newcommand{\ed}{\end{document}}
\newcommand{\bc}{\begin{center}}
\newcommand{\ec}{\end{center}}
\newcommand{\bfr}{\begin{flushright}}
\newcommand{\efr}{\end{flushright}}
\newcommand{\lt}{\left}
\newcommand{\rt}{\right}
\newcommand{\vs}{\vspace}
\newcommand{\hs}{\hspace}
\newcommand{\beq}{\begin{equation}}
\newcommand{\eeq}{\end{equation}}
\newcommand{\lb}{\linebreak}
\newcommand{\pb}{\pagebreak}
\newcommand{\mb}{\makebox}
\newcommand{\fb}{\framebox}
\newcommand{\mc}{\multicolumn}
\newcommand{\ben}{\begin{enumerate}}
\newcommand{\een}{\end{enumerate}}
\newcommand{\bit}{\begin{itemize}}
\newcommand{\eit}{\end{itemize}}
\newcommand{\ol}{\overline}
\newcommand{\un}{\underline}
\newcommand{\lefq}{\lefteqn}
\newcommand{\ba}{\begin{array}}
\newcommand{\ea}{\end{array}}
\newcommand{\beqa}{\begin{eqnarray}}
\newcommand{\eeqa}{\end{eqnarray}}
\newcommand{\beqas}{\begin{eqnarray*}}
\newcommand{\eeqas}{\end{eqnarray*}}
\newcommand{\bfg}{\begin{figure}}
\newcommand{\efg}{\end{figure}}
\newcommand{\bds}{\begin{displaymath}}
\newcommand{\eds}{\end{displaymath}}
\newcommand{\btb}{\begin{tabbing}}
\newcommand{\etb}{\end{tabbing}}
\bc {
\textbf{\large \textbf{ THE EXTENSION OF THE MASSLESS FERMION IN THE COSMIC STRING  SPACETIME} } } \\
\vspace*{1.0cm}
\large Beng\"{u} \c{C}a\~{g}atay$^{a}$, \"{O}zlem Ye\c{s}ilta\c{s}$^{b}$, An{\i}l L. Ayg\"{u}n$^{b}$ \\
\vspace*{.5cm}
\small $^a$ Turkish Atomic Energy Authority,
Sarayk\"{o}y Nuclear Research and Training Centre, 06983 Ankara, Turkey, e-mail: bengudemircioglu@yahoo.com  \\
\small $^b$ Gazi University, Faculty of Science, Physics Department 06500 Teknikokullar/Ankara Turkey, e-mail: yesiltas@gazi.edu.tr
\end{center}
\noindent \\

\vspace*{.5cm}
\begin{abstract}
\noindent In this work, we have obtained the solutions of a massless fermion which is under the external magnetic field around a cosmic string  for specific three potential models using supersymmetric quantum mechanics. The constant magnetic field, energy dependent potentials and position dependent mass models are investigated for the Dirac Hamiltonians and an extension of these three potential models and their solutions are also obtained. The energy spectrum and potential graphs for each case are discussed for the $\alpha$ deficit angle.

\end{abstract}
\noindent {\bf keyword:}   Cosmic strings, Dirac equation, curved space-time, graphene \\

\section{Introduction}
The massless Dirac character of the low energy electrons moving has attracted much interest in physics due to the graphene's  important electronic properties \cite{1}. There are  a series of studies on the interaction of graphene electrons in perpendicular magnetic fields  have been carried out in order to find a way for confining the charges \cite{2}, \cite{3}. Graphene and it's dervitaves (nanotubes, fullerenes) have become a mine of novel technologies and  studying various aspects of physics  through it's many subfields as well as including cosmological models on honeycomb branes \cite{4}. Moreover, toplogical defects which are disorders lead to important effects on the electronic properties of low dimensional systems. Cosmic string and topological defect relationship is studied where the  topological defects located at arbitrary positions on the graphene plane \cite{5}. In line with the efforts to combine general relativity theory with quantum mechanics, these studies are of great interest \cite{6}, \cite{7}, \cite{8}. Cosmic strings were introduced by Kibble in 1976 \cite{9} and the geometry of the massless cosmic strings are  examined in the large scale limit of the model given in \cite{10}. In relativistic quantum mechanics, Lie algebraic approaches \cite{11}, a hydrogen atom in the background of an infinitely thin cosmic string \cite{12}, scalar particle dynamics  in gravity's rainbow through the space-time of a cosmic string \cite{13}, $N=2$ supersymmetric approach to cosmic string dynamics \cite{14} are some recent studies about the topic. The origin of Supersymmetric(SUSY) quantum mechanics is based on very early in the development of quantum mechanics Dirac found a method to factorize the harmonic oscillator and Schrodinger noted symmetries in the solutions to his equation\cite{15}. After then, in the context of  SUSY QM was first studied  by Witten \cite{16} and Cooper and Freedman \cite{17}. This theory, which still attracts much attention today, has also application in optics \cite{18}, biophysics \cite{19}, it it has a very important place in both relativistic \cite{20}  and non-relativistic quantum mechanics \cite{21}, \cite{22}. This work points the dynamics of a massless fermion out when it is under the influence of magnetic field. Using the fundamental aspects of SUSY quantum mechanics, three different potential models are discussed for the different values of deficit angle. Moreover, supersymmetric extension of these models provides generating unknown and complex potential models. The paper is organized as follows: low dimensional Dirac equation is written in the curved spacetime using cosmic string line element and aspects of SUSY quantum mechanics are given in Section II. Section III is devoted to all potential models which are radial, hyperbolic and nonlinear ones and their solutions are given. Their energy and potential function graphes are shown. Section IV includes the extended new quantum mechanical potential models and their solutions. Energy and potential graphs are also shown. We conclude our results in Section V.
\section{Dirac Hamiltonian for a cosmic string in the gravitational background}
A massless fermion dynamics can be respresented by the Dirac equation-Weyl equation. Specially, we assume that the particle occures in the external electromagnetic field in a cosmic string spacetime. Then, the particle is described by
\begin{equation}\label{1}
  i\sigma^{a}e^{\mu}_{a}(\nabla_{\mu}+ieA_{\mu})\psi=0.
\end{equation}
Here $\sigma^{a}$ are the Pauli matrices, $\nabla_{\mu}=\partial_{\mu}+\Gamma_{\mu}$ is the covariant derivative. The cosmic string spacetime is represented by the line element which is
\begin{equation}\label{2}
  ds^{2}=-dt^{2}+dr^{2}+\alpha^{2}r^{2}d\phi^{2}+dz^{2}
\end{equation}
$-\infty <(t,z) <\infty$, $r\geq 0$, $0\leq \phi \leq 2\pi$. The parameter $\alpha=1-\frac{4\tilde{m}}{c^{2}}$ is the angular deficit changing in the interval $(0,1]$, $\tilde{m}$ is the linear mass density of the cosmic string. Here, $e^{a}_{\mu}$ are called as the tetrad fields which connect the Riemannian metric tensor the the flat spacetime metric tensor as
\begin{equation}\label{3}
  g^{\mu \nu}=e_{a}^{\mu}e_{b}^{\nu}\eta^{ab}.
\end{equation}
The tetrads and spinor connections are given as respectively
\begin{equation}\label{a4}
  e^{a}_{\mu}(x)=\begin{pmatrix}
                   1 & 0 & 0 & 0 \\
                   0 & \cos \phi & -\alpha r \sin \phi & 0 \\
                   0 & \sin \phi & \alpha r \cos \phi & 0 \\
                   0 & 0 & 0 & 1 \\
                 \end{pmatrix}
\end{equation}
\begin{equation}\label{a5}
  \Gamma_{\mu}=\frac{i}{4}\omega_{\mu a b }\Sigma^{ab}
\end{equation}
where
\begin{equation}\label{a6}
  \Sigma^{ab}=\frac{i}{2}[\gamma^{a}, \gamma^{b}]
\end{equation}
where $\gamma^{a}$ the standard Dirac matrices defined in the Minkowski spacetime:
\begin{equation}\label{a7}
  \gamma^{0}=\beta=\left(
                     \begin{array}{cc}
                       1 & 0 \\
                       0 & -1 \\
                     \end{array}
                   \right),  \gamma^{i}=\beta \alpha^{i}=\left(
                                                           \begin{array}{cc}
                                                             0 & \sigma^{i} \\
                                                             -\sigma^{i} & 0 \\
                                                           \end{array}
                                                         \right), \Sigma^{i}=\left(
                                                                               \begin{array}{cc}
                                                                                 \sigma^{i} & 0 \\
                                                                                 0 & \sigma^{i} \\
                                                                               \end{array}
                                                                             \right).
\end{equation}
We use the units $\hbar=c=1$, the line element of the stationary cosmic string spacetime is written as
\begin{equation}\label{a8}
  ds^{2}=-dt^{2}+dr^{2}+\alpha^{2}r^{2}d\phi^{2}+dz^{2}
\end{equation}
with $-\infty<(t,z)<\infty$ and $r \geq 0$, $0\leq  \phi \leq 2\pi$. Here, the parameter $\alpha$ is the angular deficit changing in the interval $(0, 1]$. Considering the components of the metric tensor $g_{\mu \nu}$,
\begin{equation}\label{a9}
  g_{11}=-1, ~~g_{22}=1, ~~g_{33}=\alpha^{2}r^{2},~~g_{33}=1,
\end{equation}
and the Christoffel symbols
\begin{equation}\label{a10}
  \Gamma^{\mu}_{ij}=\frac{1}{2}g^{\mu k}\left(\frac{\partial g_{ik}}{\partial q^{j}}+\frac{\partial g_{jk}}{\partial q^{i}}-\frac{\partial g_{ij}}{\partial q^{k}}\right),
\end{equation}
the spin connection components can be calculated using the tetrad components and the Christoffel symbols,
\begin{equation}\label{a11}
  \omega^{a}_{\mu b}=\eta_{ac}e^{c}_{\nu} e^{\sigma}_{b}\Gamma_{\sigma \mu}^{\nu}-\eta_{ac}e_{b}^{\nu}\partial_{\mu}e_{\nu}^{c}
\end{equation}
which leads to
\begin{equation}\label{a12}
  \omega_{\phi ab}=\left(
                     \begin{array}{cccc}
                       0 & 0 & 0 & 0 \\
                       0 & 0 & 1-\alpha & 0 \\
                       0 & -(1-\alpha) & 0 & 0 \\
                       0 & 0 & 0 & 0 \\
                     \end{array}
                   \right).
\end{equation}
Substituting (\ref{a12}), (\ref{a7}), (\ref{a9}), (\ref{a4}) in (\ref{1}), we can obtain \cite{hos}
\begin{equation}\label{a13}
\begin{split}
  i \gamma^{0}\frac{\partial \Psi}{\partial t}+i\gamma^{1}(\cos \phi \frac{\partial}{\partial r}-\frac{\sin \phi}{\alpha r}\frac{\partial}{\partial \phi}+\frac{i \sin \phi}{\alpha r}\frac{\omega^{1}_{\phi 2}}{2}\gamma^{3}-\frac{i\sin\phi}{\alpha r}e A_{\phi})\Psi+\\
  i\gamma^{2}(\sin \phi \frac{\partial}{\partial r}+\frac{\cos \phi}{\alpha r}\frac{\partial}{\partial \phi}-\frac{i \cos \phi}{\alpha r}\frac{\omega^{1}_{\phi 2}}{2}\gamma^{3}+\frac{i\cos\phi}{\alpha r}e A_{\phi})\Psi+ i\gamma^{3}\frac{\partial \Psi}{\partial z}=0.
\end{split}
\end{equation}
Let us use the following form for the spinor as
\begin{equation}\label{a14}
  \psi=\exp(-iEt+im\phi+ikz)\left(
                              \begin{array}{c}
                                \varphi_1(r) \\
                                \varphi_2(r). \\
                              \end{array}
                            \right)
\end{equation}
Putting (\ref{a14}) in (\ref{a13}) gives
\begin{equation}\label{a15}
  \left(
    \begin{array}{cc}
      E-k & i e^{-i\phi}(\frac{\partial}{\partial r}+\frac{m}{\alpha r}+\frac{1-\alpha}{2\alpha r}+\frac{e}{\alpha}\frac{A_{\phi}(r)}{r}) \\
      i e^{i\phi}(\frac{\partial}{\partial r}-\frac{m}{\alpha r}+\frac{1-\alpha}{2\alpha r}-\frac{e}{\alpha}\frac{A_{\phi}(r)}{r}) & E-k  \\
    \end{array}
  \right)\left(
           \begin{array}{c}
             \varphi_1(r) \\
             \varphi_2(r) \\
           \end{array}
         \right)
  =0.
\end{equation}
Then, we get a couple of differential equations:
\begin{equation}\label{a16}
   \left (\frac{d^{2}}{dr^{2}}-\frac{\alpha-1}{\alpha r}\frac{d}{dr}+E^{2}-k^{2}-\frac{(1+2m-3\alpha)(-1+2m+\alpha)}
  {4\alpha^{2}r^{2}}+\frac{e(\alpha-2m)A_{\phi}(r)}{r^{2}\alpha^{2}}-\frac{e^{2}A^{2}_{\phi}(r)}{r^{2}\alpha^{2}}-\frac{eA'_{\phi}(r)}{\alpha r}\right)\varphi_1(r)=0
\end{equation}
\begin{equation}\label{a17}
  \left (\frac{d^{2}}{dr^{2}}-\frac{\alpha-1}{\alpha r}\frac{d}{dr}+E^{2}-k^{2}-\frac{(1+2m-\alpha)(1-2m-3\alpha)}
  {4\alpha^{2}r^{2}}-\frac{e(\alpha+2m)A_{\phi}(r)}{r^{2}\alpha^{2}}-\frac{e^{2}A^{2}_{\phi}(r)}{r^{2}\alpha^{2}}+\frac{eA'_{\phi}(r)}{\alpha r}\right)\varphi_2=0
\end{equation}
Next, we transform the system given above into the form
\begin{eqnarray}\label{a18}
   \left(-\frac{d^{2}}{dr^{2}}+W(r)^{2}+W'(r)\right)\chi_1(r)&=& \varepsilon \chi_{1}(r)  \\ \label{a19}
   \left(-\frac{d^{2}}{dr^{2}}+W(r)^{2}-W'(r)\right)\chi_2(r) &=&   \varepsilon \chi_{2}(r)
\end{eqnarray}
where $\varepsilon=E^{2}-k^{2}$ and
\begin{eqnarray}\label{a20}
  \chi_{1,2}(r) &=& r^{\frac{1}{2}(\frac{1}{\alpha}-1)} \varphi_{1,2}(r)\\ \label{a21}
  W(r) &=& \frac{eA_{\phi}(r)+m}{\alpha r}
\end{eqnarray}
are used. Here, (\ref{a16}) and (\ref{a17}) are transformed into (\ref{a18}) and (\ref{a19}) with the same energy which also shows that the system is supersymmetric. Let us call two effective Hamiltonians for (\ref{a18}) and (\ref{a19}) as $H_1$ and $H_2$ respectively:
\begin{equation}\label{22}
  H_1=-\frac{d^{2}}{dr^{2}}+V_1(r)=\mathcal{L}^{-}\mathcal{L}^{+},~~H_2=-\frac{d^{2}}{dr^{2}}+V_2(r)=\mathcal{L}^{+}\mathcal{L}^{-}
\end{equation}
and the intertwining operators are defined as
\begin{equation}\label{23}
  \mathcal{L}^{\pm}=\mp\frac{d}{dr}+W(r).
\end{equation}
It is noted that (\ref{23}) can be used to interwine the system as
\begin{equation}\label{230}
  H_1 \mathcal{L}^{-}=\mathcal{L}^{-}H_2,~~~~H_2 \mathcal{L}^{+}=\mathcal{L}^{+}H_1.
\end{equation}
Furthermore, one can observe that
\begin{equation}\label{24}
  \mathcal{L}^{-}\chi_{2}=\epsilon \chi_1,~~~~\mathcal{L}^{+}\chi_{1}=\epsilon \chi_2.
\end{equation}
Let us discuss the exactly solvable potential models for $H_1$ and $H_2$ in the next Section.
\section{Potential Models}
Now we can arrange some different $A_{\phi}(r)$ vector potential models which  give rise to effective Hamiltonians. We can argue that the discrete spectrum of the Hamiltonian $H_2$ is $\epsilon_{2,n}$ and $\mathcal{L}^{-}\chi_{2,0}=0$, then,   \cite{SK}
\begin{equation}\label{25}
  \epsilon_{1,n-1}=\epsilon_{2,n}
\end{equation}
\begin{equation}\label{26}
  \chi_{1,n-1}(r)=\frac{1}{\sqrt{\epsilon_{2,n}}}\mathcal{L}^{-}\chi_{2,n}(r),~~~~n=1,2,...
\end{equation}
\subsection{Constant Magnetic Field}

The constant magnetic field vector $\vec{B}=[0,0,a_{0}]$, $a_{0}$ is a real parameter, is perpendicular to the plane. Then, vector potential component $A_{\phi}(r)$ can be taken as $A_{\phi}(r)=a_{0}r$ and $W(r)$ becomes
\begin{equation}\label{27}
  W(r)=\frac{e a_{0}}{\alpha}+\frac{m}{\alpha r}.
\end{equation}
Here we have considered the one dimensional system. Hence, $V_{1,2}(r)$ functions become
\begin{eqnarray}\label{28}
  V_1(r) &=& \frac{a^{2}_{0}e^{2}}{\alpha^{2}}+\frac{m(m-\alpha)}{\alpha^{2}r^{2}}+\frac{2a_{0}e m}{\alpha^{2}r} \\
  V_2(r) &=&  \frac{a^{2}_{0}e^{2}}{\alpha^{2}}+\frac{m(m+\alpha)}{\alpha^{2}r^{2}}+\frac{2a_{0}e m}{\alpha^{2}r}
\end{eqnarray}
This system shows that $a_0 $ should be negative because of the Coulomb's potential, let $a_0$ be $a_0=-\nu, \nu >0$. The solutions of the system (\ref{28}) which is known as pseudoharmonic potential are already known \cite{RS}:
\begin{equation}\label{29}
  E_{n}=\pm \sqrt{k^{2}-\frac{\nu^{2}e^{2}}{\alpha^{2}}-\frac{a_0^{2} e^{2} m^{2}}{\alpha^{4}\left(n+\frac{1}{2}+
  \sqrt{\frac{m(m-\alpha)}{\alpha^{2}}+\frac{1}{4}}\right)^{2}}},~~n=0,1,2,...
\end{equation}
\begin{equation}\label{30}
  \chi_{1,n}=N_{n}r^{\frac{1}{2}+\mu} \exp(-\epsilon r)~~_{1}F_{1}(-n,2\mu+1,2\epsilon r)
  \end{equation}
where $\mu=\frac{m(m-\alpha)}{\alpha^{2}}+\frac{1}{4}$, $\epsilon=\frac{\nu^{2}e^{2}}{\alpha^{2}}-E $, $_{1}F_{1}(-n,2\mu+1,2\epsilon r)$ are the confluent hypergeometric functions. The normalization constant can be written as \cite{RS}
\begin{equation}\label{31}
  N_n=\Gamma(2\mu+1)\sqrt{\frac{n! (2n+2\mu+1)}{\Gamma(n+2\mu+1)}}.
\end{equation}
Let's see how the energy and potential functions change with $\alpha$.
\begin{figure}[htbp]
\begin{center}
\caption{The graph of (\ref{28}) and (\ref{34}). $V_1(r)$ graph versus position for the different values $\alpha$. $\alpha=0.5$ for orange, $\alpha=0.5$ for blue, $\alpha=0.8$ for green curves and the red dashed curve stands for the vector potential. }
\includegraphics{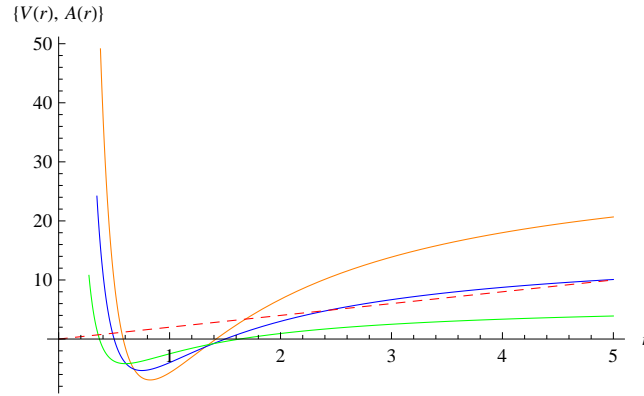}
\label{fig3}
\end{center}
\end{figure}

\begin{figure}[htbp]
\begin{center}
\caption{The graph of energy eigenvalues with respect to $\alpha$ (\ref{29}). The green curve is drawn for $k=10, n=1$, $a_0=2$, the yellow curve is shown for $k=8, n=2$ and $a_0=0.2$. }
\includegraphics{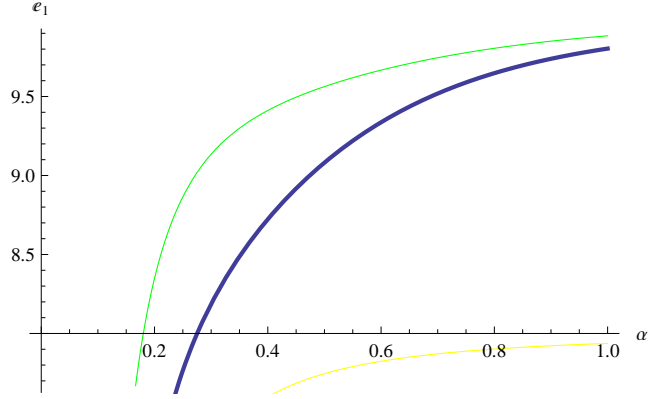}
\label{fig3}
\end{center}
\end{figure}
As is shown from Figure 1, we can obtain physical potentials for the values of $\alpha$ which takes $0 < \alpha< 1$. Energy graph in Figure 2 shows that for the values of $\alpha\geq 0.1831$ (green curve) we can get an increasing energy graph. For the greater values of the number $n$, the lower bound for the $\alpha$ increases.
\newpage
\subsection{Potential Models: Energy dependent vector and scalar potentials}
The energy dependent potentials are one of the some modified problems of the quantum mechanics \cite{L}. The condition on the density distribution is \cite{L}
\begin{equation}\label{32}
  \rho_{n}(E_n,r)=|\psi_n(E_n,x)|^{2}(1-\frac{\partial V(E_n,r)}{\partial E_n}).
\end{equation}
As it is seen from the above equation, a positivity condition is
\begin{equation}\label{33}
  1-\frac{\partial V(E_n,r)}{\partial E_n} >0.
\end{equation}
The details of the related energy dependent potentials in supersymmetry can be found in \cite{L1}. In this case we take the vector potential $A(r)$ as a complex function
\begin{equation}\label{34}
  A(r)=(C_1+S(r))r-\frac{m}{e}
\end{equation}
where $C_1$ is a constant, $S(r)$ is the complex function. Then, (\ref{a18}) and (\ref{a19}) become
\begin{equation}\label{35}
  \chi''_{1}(r)+\left(E^{2}-k^{2}-\frac{C^{2}_{1}e^{2}}{\alpha^{2}}-\frac{2C_1 e^{2}S(r)}{\alpha^{2}}-\frac{e^{2}S(r)^{2}}{\alpha^{2}}-\frac{e}{\alpha}S'(r)\right)\chi_{1}(r)=0,
\end{equation}
\begin{equation}\label{36}
  \chi''_{2}(r)+\left(E^{2}-k^{2}-\frac{C^{2}_{1}e^{2}}{\alpha^{2}}-\frac{2C_1 e^{2}S(r)}{\alpha^{2}}-\frac{e^{2}S(r)^{2}}{\alpha^{2}}+\frac{e}{\alpha}S'(r)\right)\chi_{2}(r)=0,
\end{equation}
where we can give the superpotential $W(r)$ as
\begin{equation}\label{37}
  W(r)=\frac{C_1 e}{\alpha}+\frac{e S(r)}{\alpha}.
\end{equation}
and let $S(r)$ be
\begin{equation}\label{38}
  S(r)=A_1 \sec h r+B_1 \tanh r-C_1,
  \end{equation}
where $A_1, B_1, C_1$ are the constants. We can get the partner potentials as
\begin{eqnarray}
  V_1((r) &=&\frac{B^{2}_{1}e^{2}}{\alpha^{2}}+\frac{e(A^{2}_{1}e+B_{1}(-B_{1}e+\alpha))}{\alpha^{2}}\sec h^{2} r+\frac{A_1e(2B_{1}e-\alpha)}{\alpha^{2}} \sec h r \tanh r  \\
  V_2(r) &=&  \frac{B^{2}_{1}e^{2}}{\alpha^{2}}+\frac{e(A^{2}_{1}e-B_{1}(B_{1}e+\alpha))}{\alpha^{2}}\sec h^{2} r+\frac{A_1e(2B_{1}e+\alpha)}{\alpha^{2}} \sec h r \tanh r
\end{eqnarray}
Comparing our system (\ref{35}), (\ref{37}) and (\ref{38}) with the linear energy-dependence results in \cite{a} can give a solvable model. We can use the parameter $A_1$ as energy dependent in our calculations as
\begin{equation}\label{39}
  A_1=\frac{\alpha}{2e}\left(i-2\sqrt{E^{2}-\frac{B^{2}_{1}e^{2}}{\alpha^{2}}}\right).
\end{equation}
Then, we  match our system with the one where linear energy dependency can be found in page 9 \cite{a} and write the partner potentials as energy dependent potentials as
\begin{equation}\label{40}
  V_{1}(r)=\frac{B^{2}_{1}e^{2}}{\alpha^{2}}+\left(E^{2}+\frac{B_1 e}{\alpha}-\frac{1}{4}-\frac{2B^{2}_{1}e^{2}}{\alpha^{2}}-i\sqrt{E^{2}-\frac{B^{2}_{1}e^{2}}{\alpha^{2}}}\right)\sec h^{2}r+\frac{(\alpha-2B_{1}e)}{2\alpha}\left(-i+2\sqrt{E^{2}-\frac{B^{2}_{1}e^{2}}{\alpha^{2}}}\right)\sec h r \tanh r
\end{equation}

\begin{equation}\label{41}
  V_{2}(r)=\frac{B^{2}_{1}e^{2}}{\alpha^{2}}+\left(E^{2}-\frac{B_1 e}{\alpha}-\frac{1}{4}-\frac{2B^{2}_{1}e^{2}}{\alpha^{2}}-i\sqrt{E^{2}-\frac{B^{2}_{1}e^{2}}{\alpha^{2}}}\right)\sec h^{2}r+\frac{(\alpha+2B_{1}e)}{2\alpha}\left(i-2\sqrt{E^{2}-\frac{B^{2}_{1}e^{2}}{\alpha^{2}}}\right)\sec h r \tanh r
\end{equation}
Then, one can find the energy eigenvalues as
\begin{equation}\label{42}
  E_{n}=\pm \sqrt{k^{2}-\left(n+\frac{B_1 e}{\alpha}\right)^{2}+\frac{B^{2}_{1}e^{2}}{\alpha^{2}}},~~~~n=0,1,2,...
\end{equation}
and the solutions become
\begin{equation}\label{43}
\begin{split}
  \chi_{1,n}=(1-i\sinh r)^{\frac{1}{4}-\frac{1}{2}(n+1+\sqrt{k^{2}-\beta^{2}}+4i\beta)}(1+i\sinh r)^{\frac{1}{4}-\frac{1}{2}(n+1+\sqrt{k^{2}-\beta^{2}}-4i\beta)}\\
  P^{(-\frac{1}{2}\sqrt{(2+2n+2\sqrt{k^{2}-\beta^{2}}+2i\beta)^{2}}),-\frac{1}{2}\sqrt{(-2n-2\sqrt{k^{2}-\beta^{2}}+2i\beta)^{2}})}_{n}(i\sinh r)
\end{split}
\end{equation}
\begin{figure}
\begin{center}
\caption{The real and imaginary parts of the potential in (\ref{40}). Green and blue curves are corresponding to the real and imaginary parts respectively where $\alpha=0.2$}
\includegraphics{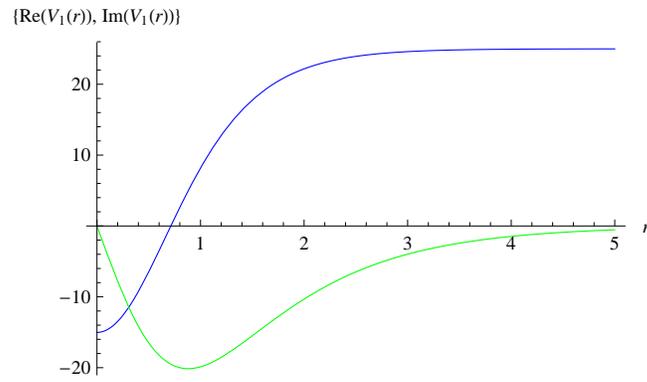}
\label{fig3}
\end{center}
\end{figure}

\begin{figure}[htbp]
\begin{center}
\caption{The real(green) and imaginary(blue) parts of the vector potential in (\ref{34}).}
\includegraphics{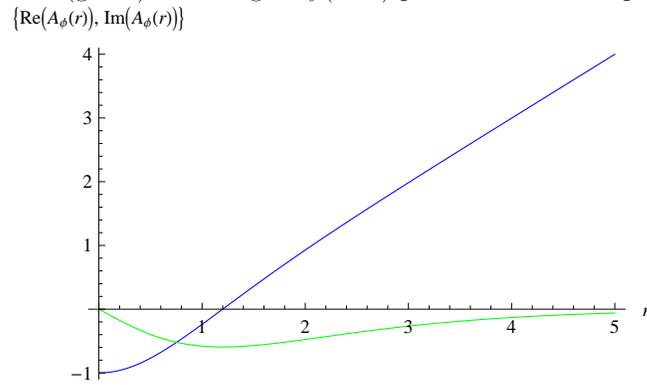}
\label{fig3}
\end{center}
\end{figure}

\begin{figure}[htbp]
\begin{center}
\caption{Energy eigenvalues verus the $\alpha$ values for (\ref{42}).}
\includegraphics{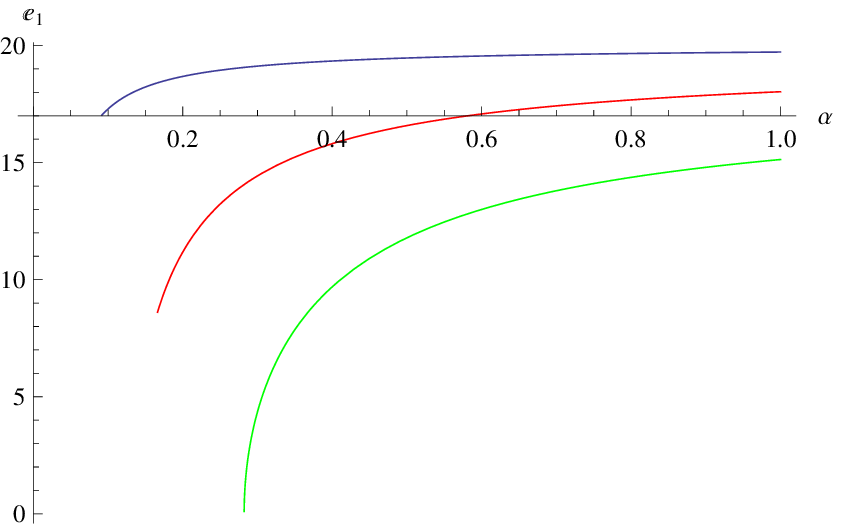}
\label{fig3}
\end{center}
\end{figure}
In Figure 3, we have obtained an effective potential graph for the real potential while we can get a potential barrier fro the imaginary part. Figure 4 shows the component of the vector potential whose real component is corresponding to another effective potential curve. In figure 5, we can see that for the greater values of $n$, the lowest bound for the $\alpha$ again increases as   energy increases.
\newpage
\subsection{Potential Models:Position dependent mass model}
Let us make a point transformation in (\ref{35}) given by
\begin{equation}\label{44}
  r=\int^{z}\frac{dt}{f(t)},
\end{equation}
then, we can get
\begin{equation}\label{45}
  f(z)^{2}\chi_{1}''(z)+f(z)f'(z)\chi_{1}'(z)+\left(E^{2}-k^{2}-\frac{C^{2}_{1}e^{2}}{\alpha^{2}}-\frac{2C_1 e^{2}S(z)}{\alpha^{2}}-\frac{e^{2}S(z)^{2}}{\alpha^{2}}+\frac{e}{\alpha}f(z)S'(z)\right)\chi_{1}(z)=0.
\end{equation}
Let us choose $f(z)=\sqrt{1+\lambda z^{2}}$ and $S(z)=\frac{iC_1}{\sqrt{1+\lambda z^{2}}}-c_1$, then, (\ref{45}) turns into
\begin{equation}\label{46}
 (1+\lambda z^{2})\chi(z)''+\lambda z \chi(z)'+\left(E^{2}-k^{2}+\frac{C^{2}_{1}e^{2}}{\alpha^{2}(1+\lambda z^{2})}+\frac{iC_1 e \lambda z}{\alpha (1+\lambda z^{2})}\right)\chi(z)=0.
\end{equation}
The solutions of the type of equations like (\ref{46}) are already known. We can write the complete solutions as \cite{midya}
\begin{equation}\label{47}
  E_{n}=\pm \sqrt{k^{2}+n\sqrt{\lambda}\left(\frac{2C_1 e-\alpha \sqrt{\lambda}}{\alpha}-n\sqrt{\lambda}\right)}
\end{equation}
and
\begin{equation}\label{48}
  \chi_{n}(z)=N_{n} i^{n}(1+\lambda z^{2})^{-s/2} e^{-r \tan^{-1}(z\sqrt{\lambda})}P^{(ir-s-1/2, -ir-s-1/2)}_{n}(iz\sqrt{\lambda}),~~~~n=0,1,2,...<s,
\end{equation}
where
\begin{equation}\label{49}
  N_{n}=\sqrt{\frac{\sqrt{\lambda}n!(s-n)\Gamma(s-ir-n+1/2)\Gamma(s+ir-n+1/2)}{\pi 2^{-2s}\Gamma(2s-n+1)}},
\end{equation}
$s=\frac{2C_1 e-\alpha\sqrt{\lambda}}{2\alpha^{3/2}}, r=-\frac{i}{2}$ and $P^{(a,b)}_{n}(y)$ are the corresponding Jacobi polynomials. Now let us examine the SUSY of this model. First, we need to make a point transformation to (\ref{46}) which is $\chi_{1}(z)=\frac{1}{(1+4\lambda^{2}z^{2})^{1/4}}Y(z)$. We get
\begin{equation}\label{049}
  Y''(z)+\left(\frac{1}{4z^{2}}+\frac{3-4iAz+4A^{2}z^{2}}{4z^{2}(1+z^{2}\lambda)^{2}}+\frac{-1+iAz+z^{2}\epsilon^{2}}{z^{2}(1+z^{2}\lambda)}\right)Y(z)=0
\end{equation}
where $A=\frac{C_1 e}{\alpha}$. We propose a super potential which is given by
\begin{equation}\label{490}
  W(z)=\frac{C_2 z^{2}}{1+z^{2}\lambda}+C_3.
\end{equation}
Then, for (\ref{049}), the unknown  parameters of the  superpotential can be obtained as
\begin{equation}\label{491}
  C_1=-\sqrt{\frac{3}{5}}\frac{2i\alpha \epsilon }{e},~~ C_2=4\sqrt{\frac{3}{5}}\epsilon^{3},~~C_3=\frac{3}{5}\epsilon,~~\lambda=-4\epsilon^{2}.
\end{equation}
\begin{figure}[htbp]
\begin{center}
\caption{Energy eigenvalues versus $\alpha$ for (\ref{47}). $ k = 1, C_1 = 1 n = 1, \lambda = 5$ for the green,  $k = 1, C_1 = 1, n = 1, \lambda = 1$ for the red and $ k = 1, C_1 = 1, n = 1, \lambda = 0.1$ for the dashed curve.}
\includegraphics{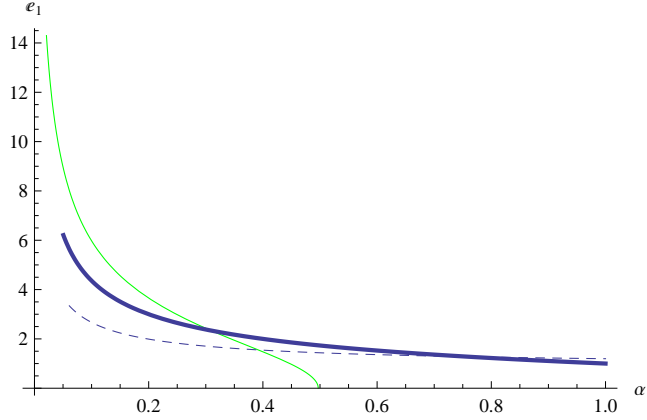}
\label{fig3}
\end{center}
\end{figure}

\begin{figure}[htbp]
\begin{center}
\caption{The effective potential graph in (\ref{46}), $\alpha=0.5$ for the dashed, $\alpha=0.9$ for the red and $\alpha=0.1$ for the magenta curves.  }
\includegraphics{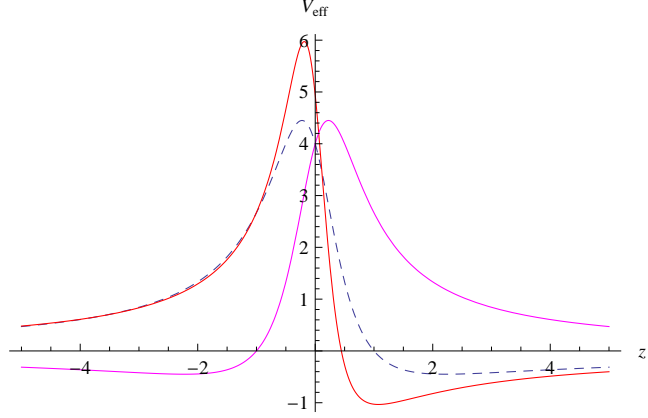}
\label{fig3}
\end{center}
\end{figure}
Figure 6 shows that the energy becomes zero when $\alpha=0.5$ and it is decreasing for the greater values of this angular deficit parameter. In Figure 7, the effective potential graph is shown where the potential well becomes more distinct as $\alpha$ increases.
\newpage

\section{Extended Potential Models }
Using $V_1$ functions given in the previous Section, we derive more general partner potentials for each model here.

\subsection{extended potential models: constant magnetic field}
Let's start with the choice of the superpotential. Here we want to generate more general potential family of the system given in (\ref{28}). Now, $W(r)$ is
\begin{equation}\label{50}
  W(r)=\frac{A}{r}+Br+\frac{f(r)}{g(r)}
\end{equation}
where $f(r), g(r)$ are unknown functions. Then, partner potentials can be obtained as
\begin{eqnarray}\label{51}
  V_1(r) &=&A+\frac{B(1+B)}{r^{2}}+\frac{2AB}{r}+\frac{f(r)^{2}}{g(r)^{2}}+\frac{2Af(r)}{g(r)}+\frac{2Bf(r)}{rg(r)}-\frac{f'(r)}{g(r)}+\frac{f(r)g'(r)}{g(r)^{2}} \\
  V_2(r) &=& A+\frac{B(-1+B)}{r^{2}}+\frac{2AB}{r}+\frac{f(r)^{2}}{g(r)^{2}}+\frac{2Af(r)}{g(r)}+\frac{2Bf(r)}{rg(r)}+\frac{f'(r)}{g(r)}-\frac{f(r)g'(r)}{g(r)^{2}} \label{52}
\end{eqnarray}
Here we match (\ref{51}) with (\ref{28}). If we equate the rational terms to zero in (\ref{51}) as below
\begin{equation}\label{52}
  \frac{f(r)^{2}}{g(r)^{2}}+\frac{2Af(r)}{g(r)}+\frac{2Bf(r)}{rg(r)}-\frac{f'(r)}{g(r)}+\frac{f(r)g'(r)}{g(r)^{2}}=0,
\end{equation}
then we obtain
\begin{equation}\label{53}
  f(r)=\frac{\exp(2Ar)r^{2B}g(r)}{C1+r^{1+2B}E_{n}(-2Ar)}
\end{equation}
where $n=-2B$ and the exponential integral function is given by
\begin{equation}\label{54}
  E_{n}(z)=\int^{\infty}_{1} \frac{\exp(-rt)}{t^{n}}dt.
\end{equation}
Then, the partner potentials become
\begin{eqnarray}\label{55}
  V_1(r) &=& A^{2}+\frac{B(B+1)}{r^{2}}+\frac{2AB}{r} \\
  V_2(r) &=& A^{2}+\frac{B(B-1)}{r^{2}}+\frac{2AB}{r}+\frac{2e^{2Ar}r^{2B-1}(e^{2Ar}r^{1+2B}+2C_1(B+Ar)+2r^{2B+1}(B+Ar)E^{-2Ar}_{n})}{(C_1+r^{2B+1}E^{-2Ar}_{n})^{2}}.
  \label{56}
\end{eqnarray}
Comparing (\ref{55}) and  (\ref{28}) gives the unkown constants $A, B$ as
\begin{equation}\label{57}
  A=-\frac{a_0 e}{\alpha}, ~~B=\frac{-m}{\alpha}.
\end{equation}
Then, one can find the solution of (\ref{56}) which is $\chi_{2,n}$ as
\begin{equation}\label{58}
  \chi_{2,n}=\frac{e^{-a_{2}r}r^{a_{1}}}{\epsilon}(c_2 n~~ _{1}F_{1}(1-n,1+c_1,c_2 r)+(A+\frac{B-a_1}{r}+a_2+\frac{e^{2Ar}r^{2B}}{C_1+r^{2B+1}E_{-2B}(-2Ar)})~~_{1}F_{1}(-n,c_1,c_2 r))
\end{equation}
where
\begin{eqnarray}\label{59}
  a_1 &=& \frac{1}{2}+\mu,~~~~\mu=\frac{m(m-\alpha1)}{\alpha^{2}} \\
  a_2 &=& \epsilon,~~c_1=2\mu+1,~~c_2=2\epsilon.
\end{eqnarray}
We note that (\ref{56}) shares the same energy level given by (\ref{29}).
\begin{figure}[htbp]
\begin{center}
\caption{$V_1(r), V_2(r)$ graphs in (\ref{55}) and (\ref{56}), $\alpha=0.3$ $a_0=-0.2$ }
\includegraphics{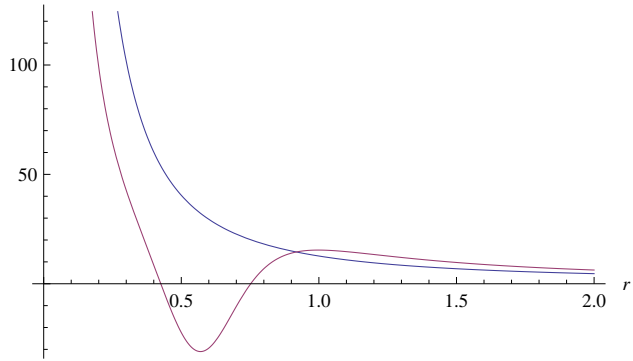}
\label{fig3}
\end{center}
\end{figure}
In Figure 8, one can see that $V_2(r)$ fits an effective potential with a well which is the partner of a Coulombic potential.
\subsection{Extended Potential Models: extended energy dependent vector and scalar potentials}
Let us choose the superpotential $W(r)$ as
\begin{equation}\label{60}
  W(r)=\frac{e(a_1+ib_1)\sec h r}{\alpha}+c_1 f(r).
\end{equation}
So, partner potentials are obtained as
\begin{eqnarray}\label{61}
  V_1(r) &=& c^{2}_{1}f(r)^{2}-c_1 f'(r)+\frac{2(a_1+ib_1)c_1 e f(r)}{\alpha}\sec h r+\frac{(a_1+ib_1)^{2}e^{2}\sec h^{2}r}{\alpha^{2}}+\frac{(a_1+ib_1)e\sec h r\tanh r}{\alpha}  \\
  V_2(r) &=&  c^{2}_{1}f(r)^{2}+c_1 f'(r)+\frac{2(a_1+ib_1)c_1 e f(r)}{\alpha}\sec h r+\frac{(a_1+ib_1)^{2}e^{2}\sec h^{2}r}{\alpha^{2}}-\frac{(a_1+ib_1)e\sec h r\tanh r}{\alpha}
\end{eqnarray}
In (\ref{60}), we can equate the terms to zero which are given below
\begin{equation}\label{62}
  c^{2}_{1}f(r)^{2}-c_1 f'(r)+\frac{2(a_1+ib_1)c_1 e f(r)}{\alpha}\sec h r=0,
\end{equation}
then, we find
\begin{equation}\label{63}
  f(r)=\frac{\exp(\frac{4e(a_1+ib_1)\tan^{-1}(\tanh \frac{r}{2})}{\alpha})(2a_1 e+2ib_1 e+i\alpha)}{2a_1 e+2ib_1 e+i\alpha-2c_1\alpha \exp(\frac{2(2a_1 e+2ib_1 e+i\alpha)\tan^{-1}(\tanh \frac{r}{2})}{\alpha})~~_{2}F_{1}\left(\frac{-2ia_1 e+2b_1e+\alpha}{2\alpha},1,\frac{3}{2}+\frac{b_1e-ia_1e}{\alpha},-e^{4i \tan^{-1}(\tanh \frac{r}{2})}\right)}.
\end{equation}
Once can give the first partner potential as
\begin{equation}\label{64}
  V_1(r)=\left(\frac{(a_1+ib_1)e}{\alpha}\right)^{2}\sec h^{2}r+\frac{(a_1+ib_1)e}{\alpha}\sec hr \tanh r
\end{equation}
and comparing the coefficients of the hyperbolic functions in (\ref{61}) with those in (\ref{40}) gives us
\begin{equation}\label{65}
  b_1=i(a_1-A_1),~~A_1=\frac{\alpha}{2e}(i-2\sqrt{E^{2}-1}),~~B_1=\frac{\alpha}{e}.
\end{equation}
Then, we can find the $V_2(r)$
\begin{equation}\label{66}
 \begin{split}
  V_2(r)&= (\frac{\bar{a}e}{\alpha})^{2}\sec h^{2}r-\frac{\bar{a}e\sec h r \tanh r}{\alpha}\\-& \frac{c_1 e^{\frac{4\bar{a}e\tan^{-1}(\tanh \frac{r}{2})}{\alpha}}(2a_1 e+i(2b_1 e+\alpha))((2a_1 e+i(2b_1e+\alpha))h(r))\sec h r}{(1+e^{4i\tan^{-1}(\tanh\frac{r}{2})}\alpha)(C_1(-2ia_1 e+2b_1 e+\alpha)+2i\alpha c_1 e^{\frac{2(2\bar{a}+i\alpha)}{\alpha}}~~_{2}F_{1}(\frac{-2ia_1e+2b_1 e+\alpha}{2\alpha},1,3/2+\frac{e(-ia_1+b_1)}{\alpha},-e^{4i\tan^{-1}(\tanh\frac{r}{2})}))^{2}}
   \end{split}
\end{equation}
where
\begin{equation}\label{67}
\begin{split}
  h(r)=(2C_1\alpha e^{\frac{2(2\bar{a}e+2i\alpha)\tan^{-1}(\tanh \frac{r}{2})}{\alpha}}+(1+e^{4i\tan^{-1}(\tanh\frac{r}{2})})(4\bar{a}eC_1+\alpha c_1 e^{\frac{4\bar{a}e\tan^{-1}(\tan h\frac{r}{2})}{\alpha}}\cosh r))\\-8\bar{a}c_1(e^{\frac{2(2\bar{a}+3i\alpha)\tan^{-1}(\tanh\frac{r}{2})}{\alpha}}+e^{\frac{2(\bar{a}+i\alpha)\tan^{-1}(\tanh\frac{r}{2})}{\alpha}})e\alpha ~~_{2}F_{1}(\frac{-2ia_1e+2b_1 e+\alpha}{2\alpha},1,3/2+\frac{e(-ia_1+b_1)}{\alpha},-e^{4i\tan^{-1}(\tanh\frac{r}{2})}).
  \end{split}
\end{equation}
and we use $\bar{a}=a_1+ib_1$, the constants $\lambda_{1,2}$ are given by

\begin{eqnarray}
  \lambda_1 &=& -(1+n+\sqrt{k^{2}-\beta^{2}}+i\beta) \\
  \lambda_2 &=&  -(-n-\sqrt{k^{2}-\beta^{2}}+i\beta)
\end{eqnarray}
\begin{figure}[htbp]
\begin{center}
\caption{$V_1(r)$ graph in (\ref{64}), $\alpha=0.3$ for the blue, $\alpha=0.9$ for the  magenta curves.$a_1=0.3, b_1=0.2$ }
\includegraphics{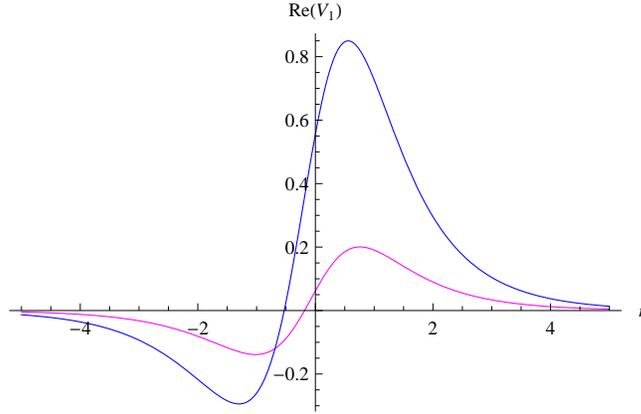}
\label{fig3}
\end{center}
\end{figure}

\begin{figure}[htbp]
\begin{center}
\caption{$V_2(r)$ in (\ref{66}), $\alpha=0.3$ for the black, $\alpha=0.9$ for the green  curves. $a_1=0.3, b_1=0.2$ }
\includegraphics{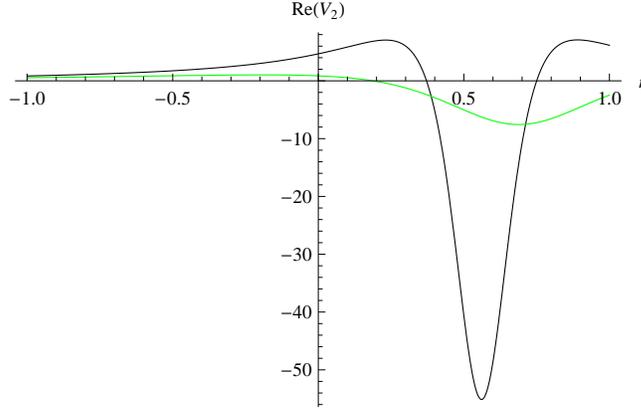}
\label{fig3}
\end{center}
\end{figure}
One can find more higher potential barrier as $\alpha$ decreases in Figure 9 for the real part of $V_1$ and for the real part of the partner potential $V_2$, it is a potential well getting more deeper as $\alpha$ decreases in Figure 10.
\newpage

\subsection{Extended Potential Models:Extended position dependent mass model}
Now we take $W(z)$ which has a form
\begin{equation}\label{69}
  W(z)=\frac{C_2 z^{2}}{1+\lambda z^{2}}+C_3+G(z),
\end{equation}
here $G(z)$ is the unknown function which can be found by the equations below
\begin{equation}\label{69}
  W(z)^{2}-W'(z) = C^{2}_{3}+\frac{C^{2}_{2}z^{4}+2C_2 z^{3}\lambda}{(1+z^{2}\lambda)^{2}}+\frac{2C_2C_3z^{2}-2C_2 z}{1+z^{2}\lambda}+2C_3 G_1(z)+\frac{2C_2 z^{2}G_1(z)}{1+z^{2}\lambda}+G_{1}(z)^{^{2}}-G_{1}'(z)
\end{equation}
\begin{equation}\label{70}
  2C_3 G_1(z)+\frac{2C_2 z^{2}G_1(z)}{1+z^{2}\lambda}+G_{1}(z)^{^{2}}-G_{1}'(z)=0.
\end{equation}
Here we note that
\begin{equation}\label{700}
  V_1(z)=\frac{\epsilon^{2}(3-8\sqrt{15}\epsilon z)}{5(1-4\epsilon^{2}z^{2})^{2}}.
\end{equation}
From (\ref{70}), $G(z)$ can be obtained as
\begin{equation}\label{71}
\begin{split}
  G(z)=\frac{2\epsilon (10+\sqrt{15})^{\sqrt{\frac{3}{5}}\tanh^{-1}(2\epsilon z)}}{-2(10+\sqrt{15})e^{\sqrt{3/5}\tanh^{-1}(2\epsilon z)}\epsilon z+2(10+\sqrt{15})\epsilon C_1 +h(z)}
  \end{split}
\end{equation}
where
\begin{equation}\label{71}
\begin{split}
  h(z)=e^{\sqrt{3/5}\tanh^{-1}(2\epsilon z)}(-(10+\sqrt{15})~~_{2}F_{1}(\frac{\sqrt{3/5}}{2},1,1+\sqrt{\frac{3}{5}}/2,-e^{2\tanh^{-1}(2\epsilon z)})+\\ \sqrt{15}e^{\sqrt{3/5}\tanh^{-1}(2\epsilon z)}~~ _{2}F_{1}(\frac{\sqrt{3/5}}{2},1,2+\sqrt{\frac{3}{5}}/2,-e^{2\tanh^{-1}(2\epsilon z)})).
\end{split}
\end{equation}
Now we can find the $V_2(z)$ as
\begin{equation}\label{72}
    V_2(z)=\frac{\epsilon^{2}}{15}\left(\frac{3(3+8\sqrt{15}\epsilon z)}{(1-4\epsilon^{2}z^{2})^{2}}\\ + c_1(z)+c_2(z)\right)
  \end{equation}
where

\begin{equation}\label{73}
  c_1(z)=\frac{600(23+4\sqrt{15})e^{2\sqrt{3/5}\tanh^{-1}(2\epsilon z)}}{(10+\sqrt{15})(-2\epsilon C_1+e^{\sqrt{3/5}\tanh^{-1}(2\epsilon z)}(2\epsilon z+_{2}F_{1}(1,\sqrt{3/2}/2,1+\sqrt{3/2}/2,1+\frac{2}{-1+2\epsilon z})))-15e^{(10+\sqrt{15})\tanh^{-1}(2\epsilon z)/5}~~\lambda(z)})^{2}
\end{equation}
\begin{equation}\label{74}
  c_2(z)= \frac{120(3+2\sqrt{15}e^{\sqrt{3/5}\tanh^{-1}(2\epsilon z)})}{(-1+4\epsilon^{2}z^{2})((10+\sqrt{15})(-2\epsilon C_1+e^{\sqrt{3/5}\tanh^{-1}(2\epsilon z)}(2\epsilon z+_{2}F_{1}(1,\sqrt{3/2}/2,1+\sqrt{3/2}/2,1+\frac{2}{-1+2\epsilon z}))-\mu(z))}
\end{equation}
and $\lambda(z)=_{2}F_{1}(1,1+\sqrt{3/5}/2+2+\sqrt{3/5}/2,1+\frac{2}{-1+2\epsilon z}$ and ~~ $\mu(z)=\sqrt{15}e^{\sqrt{3/5}\tanh^{-1}(2\epsilon z)}~~_{2}F_{1}(1,1+\sqrt{3/2},2+\sqrt{3/2},1+\frac{2}{-1+2\epsilon z})$.

\begin{figure}[htbp]
\begin{center}
\caption{The graph of the partner potentials whicch are $V_1(z)$ in (\ref{700})(blue) and $V_2(z)$ in (\ref{72}) (green)}
\includegraphics{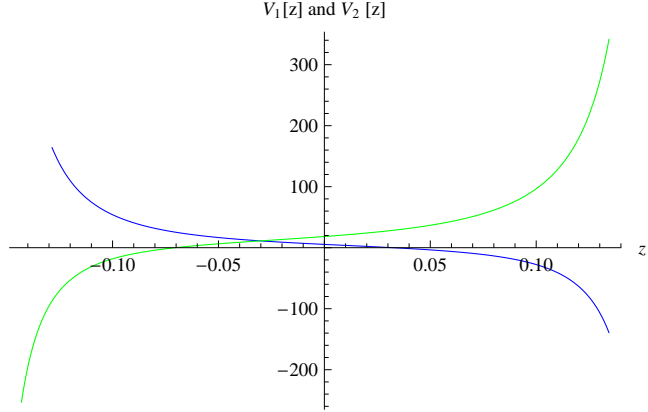}
\label{fig3}
\end{center}
\end{figure}
Figure 11 shows that the partner potentials $V_1(z), V_2(z)$ behave like $V(z)=z^{3}$ which goes to infinity as $z\rightarrow \pm \infty$. We have seen that the potential pictures can be obtained independently from $\alpha$.

\newpage

\section{Conclusions}
Using the fundamental concepts of SUSY QM, we have obtained physical solutions for the extended constant magnetic field which leads to a Coulomb problem, energy dependent hyperbolic potential and nonlinear isotonic potential which is argued as position dependent mass model for a fermion near cosmic string spacetime. It is observed that the restricted values of the angular deficit $\alpha$ gives reasonable behaviours of the potentials for our models.

\end{document}